# OPTICAL ANALOGUES OF QUANTUM CHIRALITY

D. Dragoman, A. Radu, S. Iftimie

Univ. Bucharest, Physics Dept., P.O. Box MG-11, 077125 Bucharest, Romania

**Abstract**

We show that the quantum chirality of charge carriers in graphene can be mimicked in optical structures. More precisely, we demonstrate that the particular form of the transmission coefficient at an interface in graphene between regions with different parameters can be retrieved in optics in either isotropic, gyrotropic and electro-optic materials or in complex conjugate materials. Quantitative analogies are found, at least in some parameter range, between optical and quantum parameters for which the transmission coefficient is similar in the optical and quantum cases, and hence the optical fields and quantum wavefunctions propagate in a similar manner.

**Introduction**

Quantum chirality has recently witnessed an increased interest in connection with charge carrier propagation in ballistic graphene sheets [1-2]. Unlike in common semiconductors, in which electrons and holes are described by Schrödinger equations, ballistic electron and hole states in graphene satisfy a massless Dirac-like equation. As a result, the wavefunction in graphene is spinorial, its two components expressing the contributions of the two triangular sub-lattices to the quantum states in this material. The two components of the spinorial wavefunction are related by quantum chirality, property that, as quantum entanglement, has no classical counterpart. However, classical optical analogs of locally entangled quantum states have been found and are referred to as cebits [3-4]. We show in this paper that quantum chirality can also be mimicked in classical optical systems, removing thus one fundamental barrier in the problem of establishing meaningful classical-quantum analogies.

Classical optical analogies to charge carrier propagation in graphene have been already evidenced in a limited sense, in which optical states or evolution laws have been found to be similar to their quantum counterparts [5], and even optical systems have been designed with the same transmittance/reflectance as corresponding transmission and reflection probabilities in graphene devices [6-7]. However, in all these analogies it was emphasized that the major setback for a quantitative classical-quantum analogy like that found between the classical electromagnetic field and charge carriers that obey the Schrödinger equation is that the transmission and reflection coefficients at interfaces between regions with different parameters in graphene are complex, while these coefficients are real at corresponding interfaces between optical media with no gain or loss. Therefore, classical optical systems in which the electromagnetic field propagates similarly to the chiral quantum wavefunction in graphene devices cannot be found unless we look for such a quantitative analogy in gyrotropic and electro-optic materials or in other "exotic" materials, such as the recently discovered complex conjugate media [8]. We show in this paper that, indeed, quantum chirality can be mimicked in these classical optical materials and, moreover, that a clear correspondence can be made between classical and quantum parameters such that the transmission coefficient, say, at an interface has similar values in the two cases. Similar results can be obtained if the reflection coefficient, instead of the transmission coefficient, is analyzed.

**Graphene analogies with light propagation in gyrotropic and electro-optic materials**

The time-independent ballistic propagation of charge carriers with constant energy $E$ through a graphene sheet with potential energy $V$ extending in the ($xy$) plane is described by the massless Dirac equation

$$\frac{(E-V)}{\hbar v_F}\begin{pmatrix}\psi_1\\\psi_2\end{pmatrix} = \begin{pmatrix}0 & k_x - ik_y\\ k_x + ik_y & 0\end{pmatrix}\begin{pmatrix}\psi_1\\\psi_2\end{pmatrix} \qquad (1)$$

In (1), $\psi_1$ and $\psi_2$ are the two components of the spinorial quantum wavefunction $\Psi$ in graphene and $v_F \cong c/300$ is the Fermi velocity of charge carriers that propagate with wavevector $\mathbf{k}$; $c$ is the speed of light in vacuum. Assuming that the $y$ wavevector component, $k_y$, is constant, equation (1) admits two independent plane-wave solutions with amplitudes $A$ and $B$, which represent forward- and backward-propagating waves along the $x$ direction:

$$\begin{pmatrix}\psi_1\\\psi_2\end{pmatrix} = \exp(ik_y y)\begin{pmatrix}A\exp(ik_x x) + B\exp(-ik_x x)\\ A\exp(ik_x x + i\phi) - B\exp(-ik_x x - i\phi)\end{pmatrix} \qquad (2)$$

Here $\phi = \arctan(k_y/k_x)$ and $k_x = \text{sgn}(E-V)\sqrt{(E-V)^2/(\hbar v_F)^2 - k_y^2}$ [9]. When $\text{sgn}(E-V) = 1$ electrons are the charge carriers in graphene, whereas the electrical charge is carried by holes when $\text{sgn}(E-V) = -1$.

Then, at an interface located at $x = 0$ between two regions denoted by 1 and 2, with different potential energies, $V_1$ and $V_2$, the transmission coefficient defined as $t_{gr} = A_2/A_1$ when $B_2 = 0$ is

$$t_{gr} = 2\frac{\cos\phi_1}{\exp(i\phi_2) + \exp(-i\phi_1)}; \qquad (3)$$

the energy potential in graphene can be controlled by gate electrodes [1-2]. The above expression is obtained by imposing the continuity conditions for $\psi_1$ and $\psi_2$ at the interface. The indices 1 and 2 refer to the parameter values in regions 1 and 2, respectively; for example, $A_1$ and $A_2$ are the amplitudes of the forward-propagating solution $A$ in regions 1 and 2, and $\phi_1$ and $\phi_2$ are the corresponding values of the propagation angle $\phi$. The incidence region is region 1; see also Fig. 1(a) for a graphical representation of the geometry.

In [5] it was emphasized that ballistic charge carriers in graphene propagate similarly to the electric vector **E** of a polarized electromagnetic radiation in an isotropic, gyrotropic and electro-optic material in the short-wave approximation. In this Section we explore in depth this analogy with the aim of finding a corresponding parameter in the optical case with a similar expression as the quantum transmission coefficient in (3). We do not consider in this paper the short-wave approximation, because it does not allow us to find polarization states of light with same form as the spinorial wavefunction in graphene.

As optical analog for quantum propagation in graphene we consider an isotropic, gyrotropic and electro-optic medium, in which the electromagnetic field propagates along the $z$ direction with wavenumber $\alpha$. The polarization state of a monochromatic optical field satisfies the Helmholtz equation in a non-magnetic and isotropic medium [10]:

$$(\alpha^2/k_0^2 - \varepsilon)\begin{pmatrix} E_x \\ E_y \end{pmatrix} = \begin{pmatrix} 0 & \beta - i\gamma \\ \beta + i\gamma & 0 \end{pmatrix}\begin{pmatrix} E_x \\ E_y \end{pmatrix}, \qquad (4)$$

where $\varepsilon$ is the relative permittivity of the medium, $\beta$ and $\gamma$ are the electro-optic and gyrotropic coefficients, respectively, and $k_0 = \omega/c$ with $\omega$ the frequency of the electromagnetic radiation. As for graphene, there are two independent solutions for (4), the amplitudes of which are denoted by $A$ and $B$. These solutions correspond to elliptically polarized waves with wavenumbers $\alpha_\pm = k_0\sqrt{\varepsilon \pm (\beta^2 + \gamma^2)^{1/2}}$ such that the polarization state has the form

$$\begin{pmatrix} E_x \\ E_y \end{pmatrix} = \begin{pmatrix} A\exp(i\alpha_+ z) + B\exp(i\alpha_- z) \\ \exp(i\theta)[A\exp(i\alpha_+ z) - B\exp(i\alpha_- z)] \end{pmatrix} \qquad (5)$$

with $\theta = \arctan(\gamma/\beta)$. Note that if $\beta = 0$ the two independent polarization states are the left- and right-handed circularly polarized waves [11].

Although the form of the solution (5) for the optical polarized light state is similar to (2) for the graphene wavefunction, there is a fundamental difference between them. In the quantum case, the independent solutions are plane waves propagating in opposite directions, so that their amplitudes determine the reflection and transmission coefficients at an interface between two media, whereas in the optical case the two independent polarization states propagate in the same direction, with different wavenumbers, their amplitudes being related to a change in the polarization of the electromagnetic radiation at an interface between two materials. Similarly, $\theta$ is not a propagation angle but a parameter that defines the polarization state of the electric field. In particular, at an interface located at $z = 0$ between media, labeled by 1 and 2, there is no reason to impose the condition $B_2 = 0$, unless the outgoing incident electromagnetic radiation matches the independent solution with amplitude $A_2$. This situation can occur, but, since it is easier to engineer the incident electromagnetic field, we assume in this section that $B_1 = 0$ for the incident polarized state, case in which a parameter similar to (3) can be defined as $t_{pol} = A_1 / A_2$. Then, to match the definition of the transmission coefficient for graphene, we can re-label the incident and transmission media with 2 and 1 (see also Fig. 1(b)), respectively, so that (no such re-labeling is necessary if $B_2 = 0$)

$$t_{pol} = \frac{A_2}{A_1} = 2\frac{\exp(i\theta_1)}{\exp(i\theta_1) + \exp(i\theta_2)} \tag{6}$$

Relation (6) is obtained by imposing the continuity conditions at the interface for the electric field components $E_x$ and $E_y$. Note that $t_{pol}$ does not have the significance of a transmission coefficient (in fact, the transmittance is assumed to be unity in this case), but that of the ratio of amplitudes of independent solutions in the two media. Nevertheless, this ratio can be determined from measurements of light polarization.

The condition $t_{pol} = t_{gr}$ means that both equalities: $\text{Re}(t_{pol}) = \text{Re}(t_{gr})$ and $\text{Im}(t_{pol}) = \text{Im}(t_{gr})$ must be satisfied. The first requirement, however, can be valid only if $\phi_1 = \phi_2$, i.e. if there is no electron refraction in the graphene sheet. This situation is not relevant for studying quantum-classical analogies, so that we impose instead the condition $t_{pol} = it_{gr}$, which means that the following equations must be satisfied simultaneously:

$$\frac{\sin(\theta_1 - \theta_2)}{1 + \cos(\theta_1 - \theta_2)} = \frac{\cos\phi_1(\cos\phi_1 + \cos\phi_2)}{1 + \cos(\phi_1 + \phi_2)}, \qquad (7a)$$

$$1 = \frac{\cos\phi_1(\sin\phi_1 - \sin\phi_2)}{1 + \cos(\phi_1 + \phi_2)} \qquad (7b)$$

The solution of equation (7b) is represented in Fig. 2 with red line for the case of $\text{sgn}(E - V_1) = \text{sgn}(E - V_2)$, i.e. when no electron-hole state transformation occurs at the interface, and with blue light for the situation when such a transformation occurs. The pair of propagation angles $(\phi_1, \phi_2)$ are consistent with ballistic charge carriers with energy $E$ propagating through the interface at $x = 0$ in Fig. 1, for which

$$(E - V_1)\sin\phi_1 = (E - V_2)\sin\phi_2 \qquad (8)$$

The energy $E$ determined from (8) varies as a function of $\phi_1$ as illustrated in Figs. 3(a) and 3(b) for the cases $\text{sgn}(E - V_1) = \text{sgn}(E - V_2)$ and $\text{sgn}(E - V_1) = -\text{sgn}(E - V_2)$, respectively, if $V_1 = 0.1$ eV and $V_2 = 0.4$ eV. The values of these last parameters are represented with blue lines in Figs. 3(a) and 3(b). As follows from these figures, a meaningful refraction of charge carriers in graphene, consistent with the initial simulation conditions, and $\text{sgn}(E - V_1) = -\text{sgn}(E - V_2)$, occurs only for $\phi_1 > 57°$. This is the angular interval for which an optical analog might exist, for the parameters used in our simulation.

According to (7a), a pair of variables $(\phi_1, \phi_2)$ obtained from (7b) corresponds to a shift $\Delta\theta = \theta_1 - \theta_2$ of the polarization state of the electric field. The dependence of this shift on $\phi_1$ is represented in Fig. 4, for the $\phi_1 > 57°$ range; the dependence is the same for both $\text{sgn}(E - V_1) = \pm\text{sgn}(E - V_2)$ cases. Since $\theta_{1,2} = \arctan(\gamma_{1,2} / \beta_{1,2})$ depends only on the ratio between the gyrotropic and electro-optic coefficients, an optical equivalent to a graphene interface between media with different potential energies require a material in which $\gamma$ and $\beta$ can be modified independently. In electro-optic materials $\beta$ depends on the applied electric field, while $\gamma$ can be tuned by either the electric field in optically active (or chiral) media, or the magnetic field in magneto-optic materials [12]. Therefore, we can use either different electric fields to separate the electro-optic and gyrotropic effects, taking care of their orientation with respect to the crystallographic axes of the material, or an electric and a magnetic field. We choose the latter solution in this paper and look for cadmium manganese telluride (Cd,Mn)Te single crystals as optical analogs for graphene. These crystals have both a large electro-optic and magneto-optic response, the corresponding coefficients being taken from [13-14]. Thus, for the $Cd_{0.5}Mn_{0.5}Te$ crystal, for $\theta_2 = 0$, i.e. for a non-gyrotropic materials, for example air, a $\Delta\theta = \theta_1$ shift as that in Fig. 4 can be obtained at 855 nm if the ratio between the magnetic and electric fields varies as illustrated in Fig. 5. Such a ratio can easily be applied on the $Cd_{0.5}Mn_{0.5}Te$ crystal, polarized light propagation in this material being similar to charge carrier transport in graphene.

**Graphene analogies with light propagation in complex conjugate materials**

Complex conjugate materials have been recently introduced as alternative configurations for miniaturized lasers [8]. These media are characterized by complex values of the relative electric permittivity and magnetic permeability: $\varepsilon_r = m(a+ib)$ and $\mu_r = a-ib$, respectively, such that the refractive index $n = \sqrt{\varepsilon_r \mu_r} = \sqrt{m(a^2+b^2)}$ is real. Therefore, in complex conjugate materials the electromagnetic wave propagates without loses.

In particular, for a TE electromagnetic wave incident at the interface $x = 0$ from air (medium 1) into a complex conjugate material (medium 2), the transmission coefficient has the form [8]

$$t_{ccm} = 2\frac{(a-ib)\cos\theta_1}{(a-ib)\cos\theta_1 + \sqrt{m(a^2+b^2)}\cos\theta_2}. \tag{9}$$

In this case the transmission coefficient $t = A_2 / A_1$ has the same significance as for graphene, the schematic representation in Fig. 1(c) indicating the propagation in the (*x*,*z*) plane, along the *x* direction, of the *y* component of the electric field.

As emphasized in [8], complex conjugate metamaterials can be fabricated by stacking together passive magnetic materials, with $\text{Im}[\mu_r] < 0$, and active non-magnetic dielectrics, with $\text{Im}[\varepsilon_r] > 0$. In fact, such complex conjugate materials are not uncommon in metamaterials, where they are known as zero-loss magnetic metamaterials. Such a structure, operating at MHz frequencies, consists of unit cells that incorporate a tunable phase shifter and an embedded radiofrequency amplifier [15].

To find a quantitative optical-quantum analogy we impose the condition $t_{ccm} = t_{graph}$. We have considered in our simulations a complex conjugate material with *m* = 1, *a* = 1.5 and $b = -0.05$. Then, for a given value of the $t_{ccm}$, we can find a pair of propagation angles in graphene, $(\phi_1, \phi_2)$, such that, for example,

$$\frac{\cos\phi_1 + \cos\phi_2}{\sin\phi_1 - \sin\phi_2} = \frac{\text{Re}[t_{ccm}]}{\text{Im}[t_{ccm}]} \tag{10a}$$

$$\frac{\cos\phi_1(\sin\phi_1 - \sin\phi_2)}{1 + \cos(\phi_1 + \phi_2)} = \text{Im}[t_{ccm}] \tag{10b}$$

The solution of (10a) is illustrated in Fig. 6 with blue line for $\theta_1 = 10°$ and with red line for $\theta_1 = 40°$, respectively (the two solutions are practically, indistinguishable). This pair of angles corresponds to charge carriers with an energy $E$, which varies with $\phi_1$ as illustrated in the inset of Fig. 6; the $V_{1,2}$ are the same as in the previous section and the same color as in Fig. 6 was used for the different $\theta_1$ angles.

However, if condition (10b) must be simultaneously satisfied, only one pair $(\phi_1, \phi_2)$ corresponds to a given $\theta_1$ value. This pair is found by imposing the condition $F = 0$, where $F = [\cos\phi_1(\sin\phi_1 - \sin\phi_2)]/[1 + \cos(\phi_1 + \phi_2)] - \text{Im}[t_{ccm}]$; the numerical solution of $F = 0$, and hence the corresponding $\phi_1$ to $\theta_1 = 10°$ (blue line) and $\theta_1 = 40°$ (red line) is represented in Fig. 7. Solutions of this equation have been found only for $6° < \theta_1 < 42°$, and only for $\text{sgn}(E - V_1) = \text{sgn}(E - V_2)$.

**Conclusions**

The main purpose of this paper is to demonstrate that transmission coefficients of the same form as for ballistic charge carriers at an interface between two regions in graphene with different parameters can be obtained in optics in either isotropic, gyrotropic and electro-optic materials or in complex conjugate materials. The particular form of the transmission coefficient in graphene is an expression of quantum chirality, which has no classical analog. Therefore, no optical structures with the same transmission coefficient at an interface have been found previous; only the optical transmittance at an interface was put into correspondence with the quantum transmission probability. Our results imply that quantum chirality can be mimicked in optical structures, and this result is important in its own right. It parallels the finding that local quantum entanglement can be mimicked in optical systems.

Moreover, this result is quantitative, in the sense that, at least in some parameter range, interfaces between optical media with the same transmission or reflection as their quantum counterparts can be found. Previous quantum-optical analogies involving charge carriers that obey the Schrödinger equation suggest that this condition is essential to design optical structures in which the electromagnetic field propagates similarly/has the same form as the quantum wavefunction through given quantum devices. In addition to imposing the same transmission or reflection coefficients at an interface, a phase matching condition is required [6,16] to obtain optical structures with the same transmission/reflection characteristics as a quantum device consisting of several regions with different widths, condition that determines the required width of the optical or quantum structure. Therefore, our results can be used to optically mimicked ballistic graphene devices composed of successions of regions with different parameters.

**Acknowledgements**

This work was supported by a grant of the Romanian National Authority for Scientific research, CNCS-UEFISCDI, project number PN-II-ID-PCE-2011-3-0224

**Figure captions**

Fig. 1  Schematic representation of (a) charge carrier refraction in graphene, (b) polarized light refraction in isotropic, gyrotropic and electro-optic materials, and (c) TE wave refraction at an interface between air and a complex conjugate material.

Fig. 2  Propagation angle solutions in graphene analogous to light propagation in isotropic, gyrotropic and electro-optic materials when no electron-hole state transformation occurs (red line) and when such a transformation occurs (blue line).

Fig. 3  Incidence angle dependence of the electron energy consistent with the propagation angles represented with (a) red line and (b) blue line in Fig. 2, for $V_1 = 0.1$ eV and $V_2 = 0.4$ eV (blue lines).

Fig. 4  Dependence of the shift of the optical polarization state on the correspondent incidence angle in graphene.

Fig. 5  Dependence on the incidence angle in graphene of the ratio between the magnetic and electric fields that produce the shift in Fig. 4 in a $Cd_{0.5}Mn_{0.5}Te$ crystal at 855 nm.

Fig. 6  Propagation angle solutions in graphene analogous to light propagation in a complex conjugate material illuminated at $\theta_1 = 10°$ (blue line) and $40°$ (red line). Inset: Incidence angle dependence of the electron energy consistent with the propagation angles in Fig. 6 for $\theta_1 = 10°$ (blue line) and $40°$ (red line).

Fig. 7  Numerical solution identifying through $F = 0$ the incidence angle in graphene corresponding to $\theta_1 = 10°$ (blue line) and $\theta_1 = 40°$ (red line).

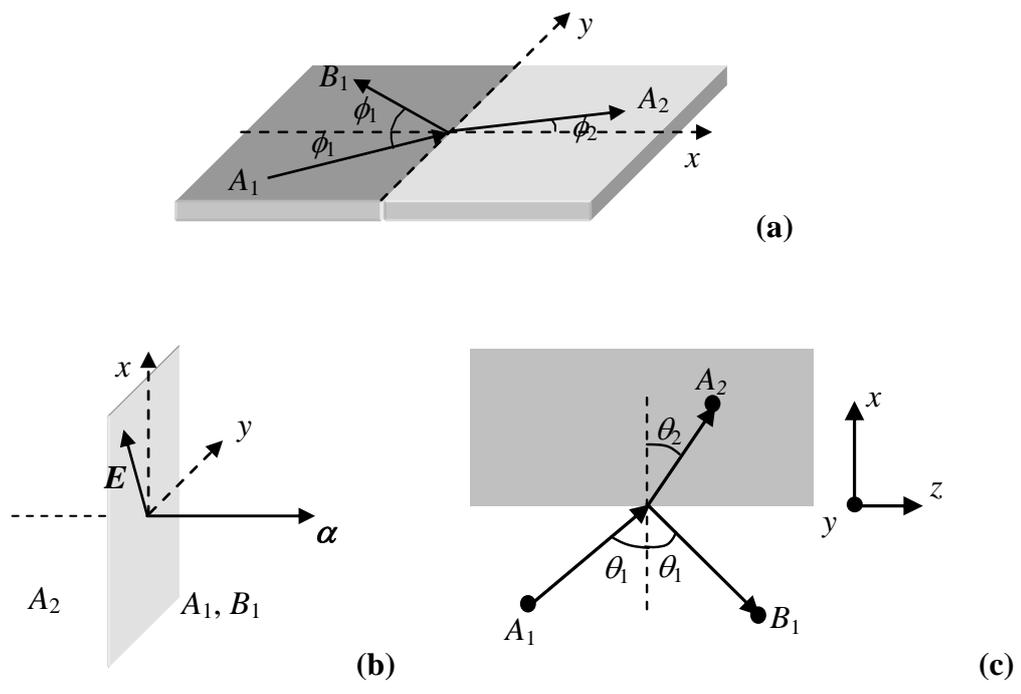

**Fig. 1**

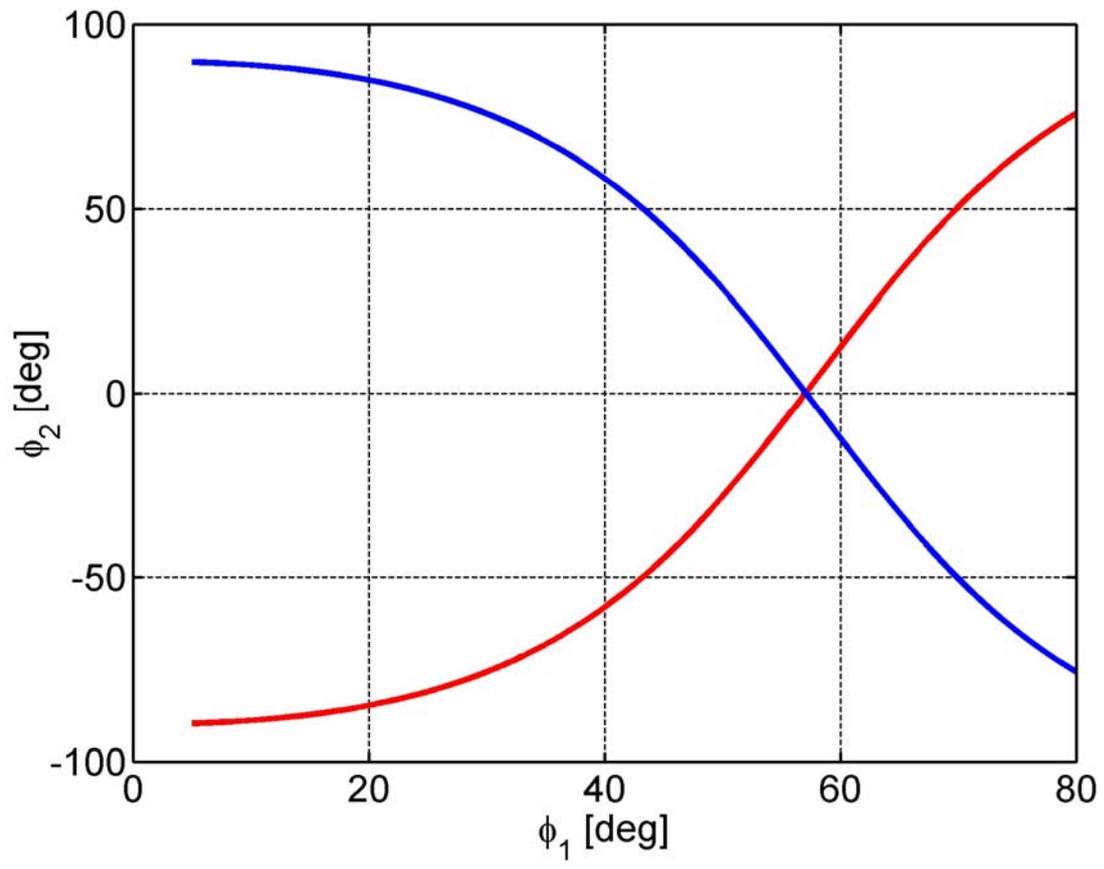

**Fig. 2**

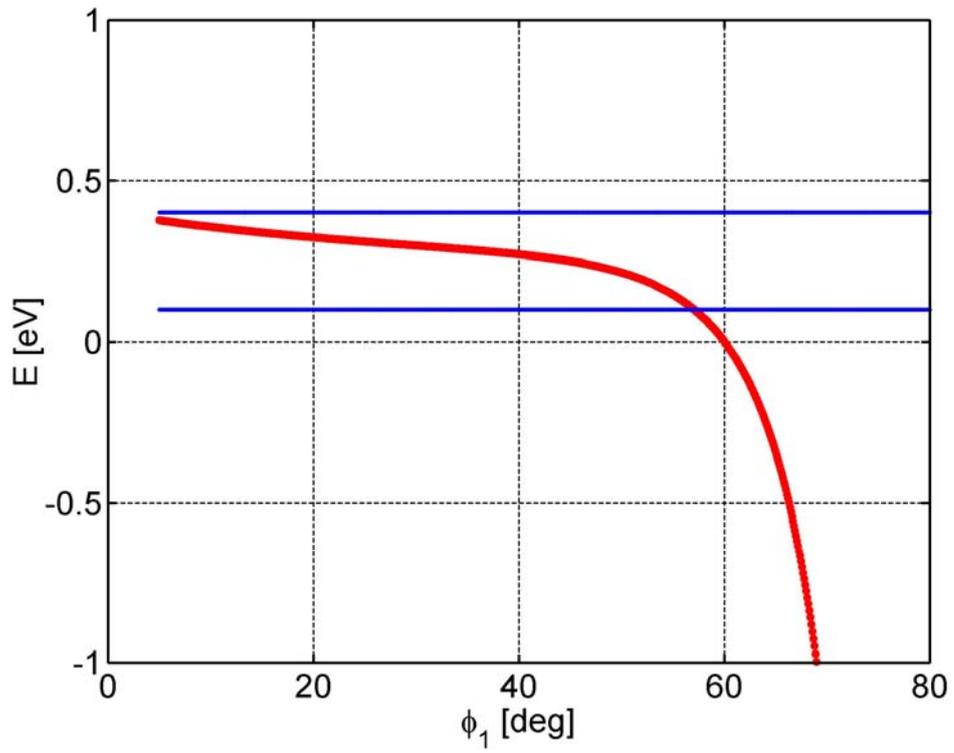

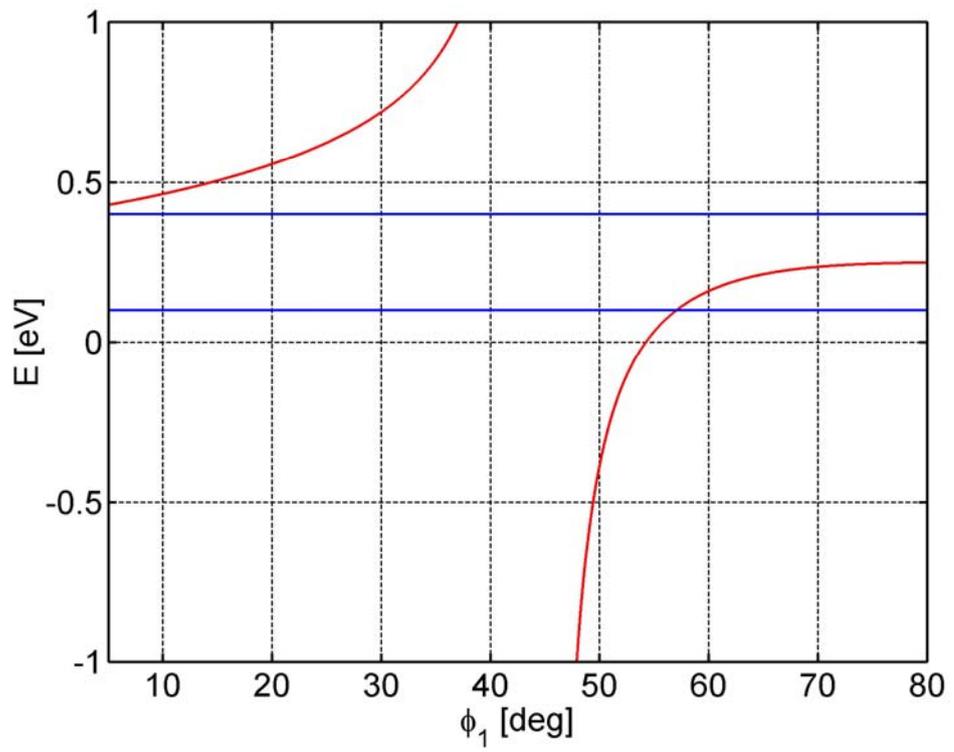

**Fig. 3**

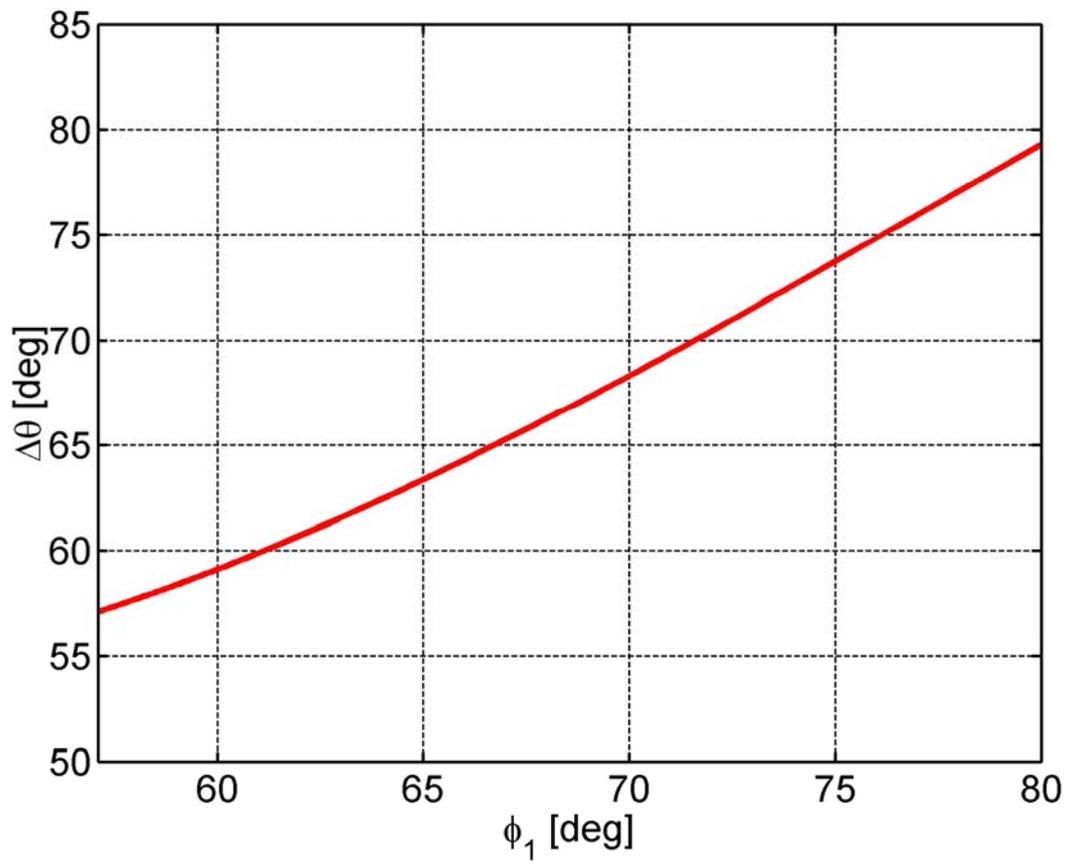

**Fig. 4**

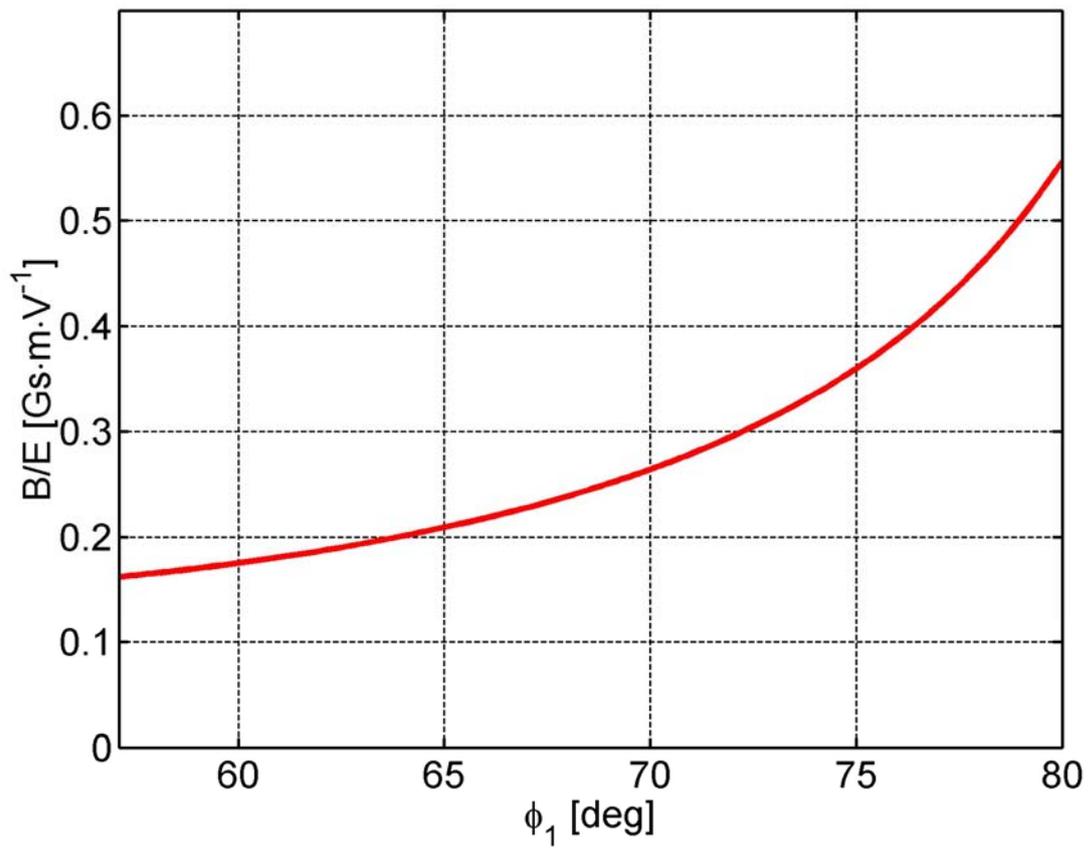

**Fig. 5**

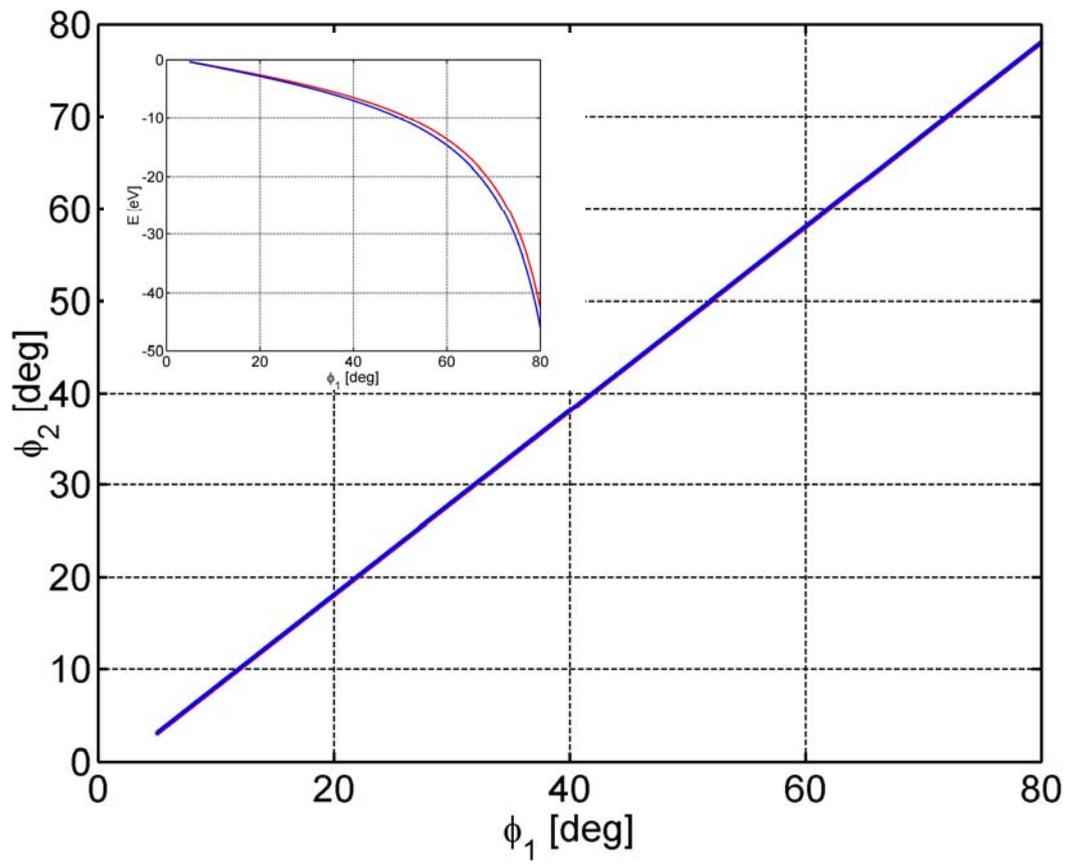

**Fig. 6**

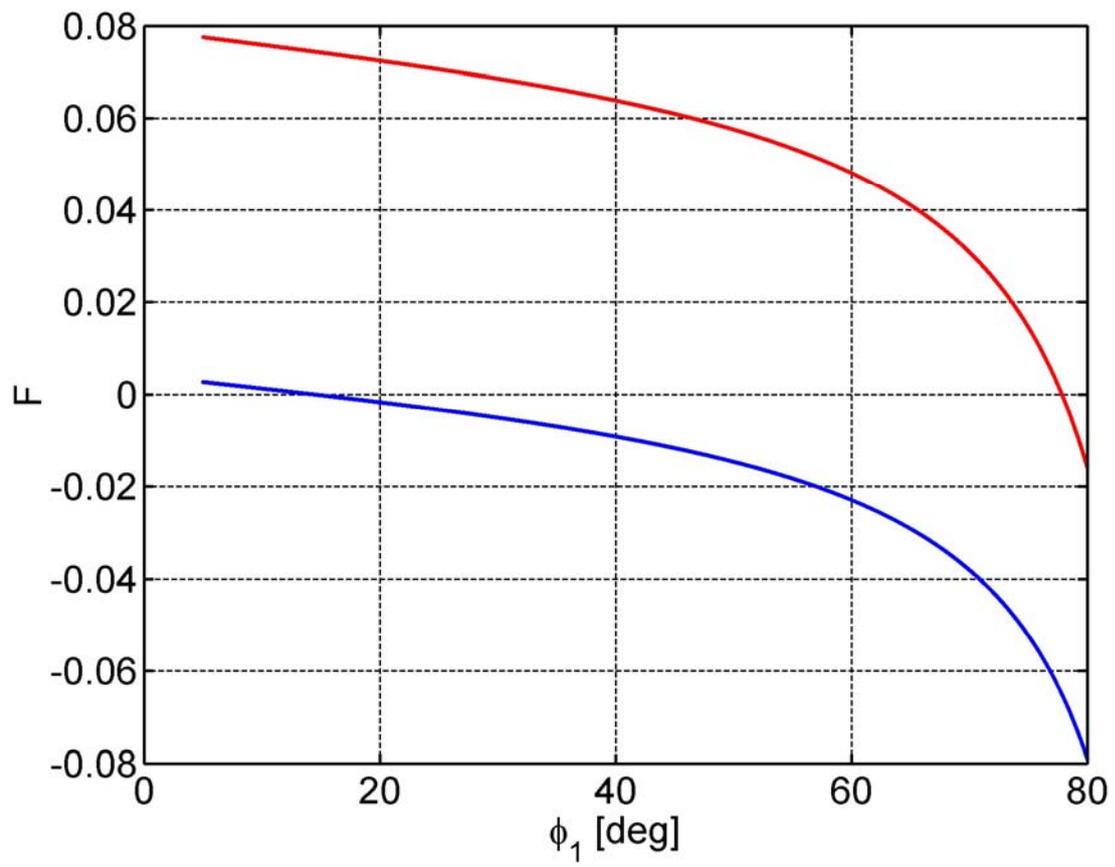

**Fig. 7**